\documentclass[12pt]{article}
\pdfoutput=1 
\usepackage{graphicx} 
\usepackage{color}      
\usepackage{cancel,tabularx,moreverb,fancybox,amsmath,float,bm,braket,slashbox,txfonts,amssymb,bm,accents,enumerate}
\usepackage[top=30truemm,bottom=30truemm,left=25truemm,right=25truemm]{geometry}
\usepackage{latexsym}
\usepackage{here}
\usepackage{cite}
\usepackage{hyperref}         

\def\be{\begin{equation}}
\def\ee{\end{equation}}
\def\ba{\begin{eqnarray}}
\def\ea{\end{eqnarray}}

 \def\ba{{\bar{\alpha}}}

  \makeatletter
    
    \@addtoreset{equation}{section}
  \makeatother

\begin{document}

\begin{titlepage}
\thispagestyle{empty}



\bigskip

\begin{center}
\noindent{{\large \textbf{
Transmission Coefficient of Super-Janus Solution
}}}\\
\vspace{2cm}
Saba Baig,
Andreas Karch,
Mianqi Wang
\vspace{1cm}

{\small \sl 
Theory Group, Weinberg Institute, Department of Physics \\ University of Texas, 2515 Speedway, Austin, TX 78712-1192, USA
}

\vskip 2em
\end{center}

\begin{abstract}
We calculate the transmission coefficient of the super-Janus interface conformal field theory, both at weak and at strong coupling, where latter is described holographically as a domain-wall solution on AdS$_2\times S^2\times M_4\times\Sigma$. Surprisingly we find perfect agreement between the free and strong coupling answer, mirroring a similar unexpected equivalence previously found for the entanglement entropy.

 \end{abstract}

\end{titlepage}

\restoregeometry

\tableofcontents

\section{Introduction}
Interface conformal field theories (ICFTs) \cite{Oshikawa:1996dj,Oshikawa:1996ww,Bachas:2001vj} share the same symmetry as the much more widely studied boundary conformal field theories (BCFTs), but display several novel features, which only recently have found wide appreciation both in the high energy and condensed matter communities.

While ICFTs can always be mapped to BCFTs via the folding trick, they give rise to novel physical quantities that, while they of course also exist in the folded picture, only make natural sense in the ICFT: the entanglement entropy (EE) of generic regions including the interface \cite{Brehm:2015lja,Sakai:2008tt,Karch:2023evr,Karch:2024udk} and the energy transmission across the interface \cite{Bachas:2001vj,Meineri:2019ycm,Quella:2006de}. In particular, in this paper the focus is on the transmission coefficient of a codimensional-1 interface dividing CFT$_1$ and CFT$_2$ with central charges $c_1$ and $c_2$, which is given by the correlation function of the stress tensor across the interface:

\begin{equation}
    \mathcal{T}=\frac{2c_{LR}}{c_1+c_2}=\frac{\langle T^1T^2+\overline{T}^1\overline{T}^2\rangle_{1|2}}{\langle (T^1+\overline{T}^1)(T^2+\overline{T}^2)\rangle_{1|2}}
\end{equation}

Together, $c_\text{eff}$ and $c_{LR}$ define two novel ``central charges" governing the properties of the interface: $c_{LR}$ appearing in the transmission, and $c_\text{eff}$ as the coefficient of log divergent terms in the EE for regions ending on the interface. Some universal bounds on these two quantities have recently been derived in \cite{Karch:2023evr,Karch:2024udk}. Another interesting quantity related to the EE is the usual boundary entropy or g-factor $g=\langle 0|B\rangle$ associated with symmetric intervals, which fold to the standard BCFT EE which is uniquely determined by $g$. For generic asymmetric intervals, $g$ actually becomes a function of $l_L/l_R$ where $l_{L/R}$ stands for the length of the part of the interval on either side of the interface. Little is known about this function $g(l_L/l_R)$. For the particular example of the bosonic Janus ICFT it has for example been worked out in \cite{Karch:2021qhd} in the strong coupling limit.

An interesting class of examples of ICFTs is provided by holographic ICFTs, that is ICFTs which in a suitable large $N$ and strong coupling limit have an equivalent description in terms of classical gravity in a higher dimensional spacetime. Simple solutions to Einstein equations that have the right symmetries to correspond to the dual of an ICFT can be constructed in terms of branes \cite{Karch:2000ct,Karch:2000gx}. Generically the field theory dual to these ``bottom-up" constructions is not known and may not even exist. To realize a concrete dual pair, we rely on ``top-down" constructions as they actually arise in string theory. The simplest such top-down ICFTs are the Janus solution and its supersymmetric generalization.

The original Janus solution \cite{Bak:2003jk} describes a 4d ICFT. The CFT on both sides of the interface is given by ${\cal N}=4$ SYM, but with different values of the coupling constant. The 2d Janus ICFT has been constructed in \cite{Bak:2007jm} and its supersymmetric generalization in \cite{Chiodaroli:2009yw}.

The 2d Janus ICFT is based on a famous holographic CFT, the D1-D5 system. 
In the free limit, the D1-D5 CFT is described by a symmetric orbifold. We start with a theory of $4N$ free compact scalars together with their free fermion superpartners. This gives a sigma model with a $(T^4)^N/S_N$ target space, which we consider in the large $N$ limit. For simplicity we are working at a special point in moduli space where $T^4$ is just a product of 4 circles with identical radii. This orbifold CFT is what we call the free limit. All properties of the orbifolded theory can be deduced from the free seed theory of free compact bosons and fermions. In order to have a holographic description, we need to drive the theory to strong coupling. Here the coupling corresponds to a marginal deformation that in terms of the sigma model is a blow up mode of the orbifold singularity. In the infinite coupling limit holography applies and the D1-D5 CFT is dual to type IIB supergravity on $AdS_3 \times S^3 \times T^4$ \cite{Maldacena:1997re}.

The (bosonic) Janus ICFT now corresponds to a deformation of the D1-D5 system where the radius of the $T^4$ jumps across the interface taking the values $r_{1,2}$ on the right and left of the interface respectively. While this jump breaks all supersymmetries, some supersymmetries can be restored by imposing non-trivial boundary conditions across the interface for the fermions as well. This latter is the super-Janus ICFT of interest to us. 

The g-factor for a symmetric interval in the bosonic Janus ICFT has been calculated at both weak and strong coupling in \cite{Azeyanagi_2008}. One should note that this is not a single number, but a function of $r_1/r_2$, the ratios of the $T^4$ radii on the two sides of the interface. While the two functions obtained in the two limits are definitely not identical, the difference is numerically remarkably small.

In contrast, in the supersymmetric case it was shown that both $g$ \cite{Chiodaroli_2010} as well as $c_\text{eff}$ \cite{Gutperle:2015hcv} as a function of $r_1/r_2$ are identical in the free theory and in the strong coupling limit, that is the CFT calculation of $S=\log g$ on the orbifold target space $T^{4N}/S_N$ matches with the holographic calculation of the RT surface \cite{Ryu:2006bv} in the bulk. While supersymmetry often gives rise to non-renormalization theorems, the remarkable insensitivity of these two entropic quantities to the coupling constant is, as of now, not understood.  

In \cite{baig2023transport} a similar comparison was performed for the energy transmission coefficient, or equivalently $c_{LR}$, for the bosonic Janus ICFT. Once again, they studied the free orbifold CFTs and its holographically dual (non-susy) Janus solution of type IIB supergravity, which is valid at strong coupling. Similar to the EE calculation, they found that without supersymmetry for a generic ratio of $r_1/r_2$, the strongly coupled transmission coefficient $\mathcal{T}$ by the holographic calculation is different, in fact strictly bigger, than that in the weakly coupling regime given by the free CFT calculation at the orbifold point. 

Given that for both entropic quantities, $c_\text{eff}$ and $g$, perfect agreement between weak and strong coupling answers had been found in the supersymmetric case, it is natural to ask whether this miracle repeats itself for the transmission coefficient. We set out to answer this question in this work by once again calculating this quantity in the free orbifold CFT, this time with a supersymmetric boundary condition, and its holographic dual valid at strong coupling. Amazingly enough we once again find perfect agreement between the weak and strong coupling answer for all values of $r_1/r_2$. All key quantities in this supersymmetric ICFT seem to be remarkably insensitive to the coupling constant of the bulk CFT, a fact which begs for a deeper understanding.

This paper is organized as follows: in Section \ref{sec:cft} we calculate the transmission coefficient in the 2d $\mathcal{N}=(4,4)$ orbifold CFT, in presence of jumping radii and $\mathbb{Z}_2$ orbifold twist operators for the free boson sector. In Section \ref{sec:holography} we calculate it in the bulk type IIB Janus model on AdS$_2\times S^2\times \Sigma$ by KK reducing the 6D supergravity to a 3D toy model, and find a perfect match with the CFT result. In Section \ref{sec:fluc} an alternative way of deriving the transmission coefficient of the super-Janus model is presented by solving the spin-2 fluctuations in AdS$_2$/CFT$_1$. Finally in Section \ref{sec:discussion} we discuss a surprising conjecture on the relation between $c_\text{eff}$ and $c_{LR}$ for supersymmetric theories, and other future directions on the subject.

\section{CFT calculation}
\label{sec:cft}

To determine the transmission coefficient in the free orbifold CFT one simply calculates the transmission coefficient for $4N$ real scalars. As shown in \cite{baig2023transport}, the fact that we are orbifolding does not affect the transport of the conserved total energy. So in the bosonic theory the task is to calculate the transmission coefficient for a free boson with compactification radius $r_1$ and $r_2$ on the two sides of the interface respectively. In fact, this system allows many different conformal interface, after folding we are looking for a BCFT of two compact scalars, or equivalently all D-branes on a torus with side lengths $r_1$ and $r_2$. As argued already in \cite{Azeyanagi_2008}, following similar arguments for the original 4d Janus ICFT in \cite{Clark:2004sb}, the particular interface boundary conditions corresponding to Janus are uniquely obtained by varying the action of the compact free bosons {\it without} adding any extra terms on the interface.

As shown in \cite{baig2023transport} these particular boundary conditions fit into the general framework of bosons whose components to the left and right of the interface are related to each other by a matrix $S$ that was analyzed already in the early work of \cite{Bachas:2001vj}. To calculate the transmission coefficient in the free Janus CFT all that needed to be done is to extract the correct form of the matrix $S$ from the boundary conditions in \cite{Azeyanagi_2008}.
Here we want to redo this calculation for the supersymmetric case. Specifically, the CFT here is a 2d $\mathcal{N}=(4,4)$ NLSM with an interface. The free-field limit again corresponds to a (supersymmetric) sigma model where the target space is an orbifold $(T^4)^N/S_N$.

One subtlety in the free theory is that the theory has more than one conserved stress tensor, so the analysis of \cite{Meineri:2019ycm} does not directly apply. In principle we could get a different transmission coefficient whether the energy comes in as fermions or as bosons: the stress tensor of free fermion and boson is separately conserved. Of course this ambiguity would disappear the moment we move away from the orbifold point which couples bosons and fermions. But we really want to calculate the transmission coefficient at the orbifold point itself. Fortunately we will see that both fermions and bosons have the same transmission coefficient in the free super-Janus ICFT, so this subtlety is immaterial. To see this note that the boundary condition for super-Janus worked out in  (6.8) of \cite{Chiodaroli_2010} once again has exactly the form of a generic free fermion interface also worked out in \cite{Bachas:2001vj}, see eq. (3.12).\footnote{To see this one has to identify in the region to the left of the interface $\Psi$ from the first reference with $\psi_-$ in the second, and $\tilde{\Psi}$ in the first with $\psi_+$ in the second, whereas in the region to the right of the interface the roles of $\psi_{\pm}$ are reversed. This flip can be attributed to the fact that the former reference works in the folded picture.}

In the comparison one finds that the angle $\vartheta$ characterizing the interface in \cite{Bachas:2001vj} is related to the radii of $T^4$ on the two sides by
\begin{equation}
\sin(2 \vartheta) = 2 \frac{r_1 r_2}{(r_1^2+r_2^2)} .
\end{equation}

This is exactly what one finds in the bosonic case: 
fermion and boson in the free super-Janus ICFT have the same transmission coefficient, which hence is equal to the transmission coefficient of total energy. That is, in free super-Janus we again get the same state-independent transmission coefficient as in the in free bosonic case:
\begin{equation}
\label{weakT}
    {\cal T}_{\text{weak}} = \frac{4}{\left ( \frac{r_1}{r_2}  +  \frac{r_2}{r_1} \right )^2}
\end{equation}

More generally, one can obtain a class of  ICFTs by adding in an exactly marginal deformation, which in turn gives the gluing condition at the boundary/interface \cite{Chiodaroli_2010}. In addition to the radii deformation above, let us consider a $\mathbb{Z}_2$ orbifold twist deformation \cite{Larsen_1999}. It is obtained by perturbing the theory with a dimension $(h,\bar{h})=(1,1)$ twist operator $T_0$, with an OPE of $T_0(z,\bar{z})T_0(w,\bar{w})\sim 1/|z-w|^4$. Under this twist, the orbifold CFTs on the two sides live on different points in their moduli space. Following \cite{Chiodaroli_2010} it was further argued that the orbifold twist deformation is one of the many possible K\"{a}hler moduli deformations. By counting distinct two forms on the orbifold point of a K\"{a}hler manifold (e.g. a $T^4/\mathbb{Z}_2$ orbifold point in $K_3$), one can naturally extrapolate the presence of orbifold twist-fields to jumping world-sheet $B$ fields across the interface. 

Concretely, let us follow Appendix C of \cite{Chiodaroli_2010} and consider two free compact bosons on each side of the interface $\sigma=0$ coupling to anti-symmetric $B$ field, with both jumping radii $r_1,r_2$ and $B$ field $b_1,b_2$ on the left and right side of the interface. The action before folding is
\begin{equation}
    \begin{split}    \int_{\sigma>0}\left(r_1^2\eta^{\mu\nu}\partial_\mu X^i\partial_\nu X^i+2\epsilon^{\mu\nu}b_1\partial_\mu X^1\partial_\nu X^2\right)   +\int_{\sigma<0}\left(r_2^2\eta^{\mu\nu}\partial_\mu \tilde{X}^i\partial_\nu \tilde{X}^i+2\epsilon^{\mu\nu}b_2\partial_\mu\tilde{X}^1\partial_\nu\tilde{X}^2\right)
    \end{split}
\end{equation}

Let us do a change of variables from $r_1$, $r_2$ and the jump in B fields $\Delta b = b_1-b_2$ to parameters $k$, $\psi$ and $\theta$:
\begin{equation}
    \Delta b=4k\sinh\theta,\quad r_1=\sqrt{\frac{2ke^{\psi}}{\cosh\psi}},\quad r_2=\sqrt{\frac{2ke^{-\psi}}{\cosh\psi}}
    \label{eq:variable}
\end{equation}
and hence in particular
\begin{equation}
\frac{r_1}{r_2} = e^{\psi}.
\end{equation}

The new parameters give a simpler expression for the transmission coefficient $\mathcal{T}$, and as we will see in Section \ref{sec:holography} they arise naturally in the bulk supergravity solution. An overall shift on the B fields will be a total derivative and leave all quantities of interest invariant.

Let us do a transformation $X^i\to r_1X^i,\tilde{X}^i\to r_2\tilde{X}^i$ as in \cite{bachas_permeable_2002} to normalize the energy-momentum tensor. The boundary conditions are
\begin{equation}
\begin{split}   
    r_1\partial_\sigma X^i-r_2\partial_\sigma \tilde{X}^i+\epsilon_{ij}\left(\frac{b_1}{r_1}\partial_\tau X^j-\frac{b_2}{r_2}\partial_\tau \tilde{X}^j\right)=0,\quad i=1,2\\
    r_2\partial_\tau X^i=r_1\partial_\tau \tilde{X}^i,\quad i=1,2\\
    \end{split}
\end{equation}

One can then write down the scattering matrix of this 2-boson system under this boundary condition at the interface, 
\begin{equation}
    \left(\begin{array}{c}
        \partial_- X^1 \\
        \partial_- X^2 \\
        \partial_+ \tilde{X}^1 \\        
        \partial_+ \tilde{X}^2 \\
    \end{array}\right) = S \left(\begin{array}{c}
        \partial_+ X^1 \\
        \partial_+ X^2 \\
        \partial_- \tilde{X}^1 \\
        \partial_- \tilde{X}^2 \\
    \end{array}\right)
\end{equation}

where we only write out the $2\times 2$ transmission submatrix $S_{21}$ for $\partial_- X^i$ and $\partial_- \tilde{X}^i$
\begin{equation}
    \left(\begin{array}{c}
        \partial_- X^1 \\
        \partial_- X^2 \\
    \end{array}\right) = \left(\begin{array}{cc}
        \frac{2r_1r_2(r_1^2+r_2^2)}{(r_1^2+r_2^2)^2+\Delta b^2}&\frac{2r_1r_2\Delta b}{(r_1^2+r_2^2)^2+\Delta b^2} \\
        \frac{2r_1r_2\Delta b}{(r_1^2+r_2^2)^2+\Delta b^2}&-\frac{2r_1r_2(r_1^2+r_2^2)}{(r_1^2+r_2^2)^2+\Delta b^2} \\
    \end{array}\right)\left(\begin{array}{c}
        \partial_- \tilde{X}^1 \\
        \partial_- \tilde{X}^2 \\
    \end{array}\right)
    \label{eq:s21}
\end{equation}

 Following \cite{Quella:2006de}, we can fold the theory via $\sigma\to-\sigma$ (and adding a minus sign in front of the $B$ field), and calculate the energy transmission coefficient using the above S matrix on the conformal boundary state $|b\rangle$. Explicitly, we can write down the $2\times 2$ $R$-matrix
 \begin{equation}
     R_{ij}=\frac{\langle 0|L_2^i\bar{L}_2^j|b\rangle}{\langle 0|b\rangle},
 \end{equation}

where $L_2^{1,2}$ are the virasoro generators for the left and right hand side.
The energy transmission coefficient is then
\begin{equation}
    \mathcal{T}_\text{weak}=\frac{2}{c_1+c_2}\left(\frac{\langle 0|L_2^1\bar{L}_2^2|b\rangle}{\langle 0|b\rangle}+\frac{\langle 0|L_2^2\bar{L}_2^1|b\rangle}{\langle 0|b\rangle}\right)=\frac{1}{2}\sum_{I,J=1,2}\big|(S_{21})_{IJ}\big|^2    
\end{equation}

Plugging in \eqref{eq:s21} leads to the result
\begin{equation}
    \mathcal{T}_\text{weak}=\frac{4}{\left ( \frac{r_1}{r_2}  +  \frac{r_2}{r_1} \right )^2+\frac{\Delta b^2}{r_1^2r_2^2}}=\frac{1}{\cosh^2\psi\cosh^2\theta}
    \label{eq:tweak}
\end{equation}

\section{Holography calculation with KK reduction}
\label{sec:holography}
In this section we consider the holographic dual of the 2D supersymmetric Janus ICFT. As mentioned above, the dual spacetime is a solution of type IIB supergravity of the form AdS$_2\times S^2\times M_4\times\Sigma$ dictated by symmetry \cite{Chiodaroli_2010}. The solution is supported by R-R and/or NS-NS charges. The manifold  $M_4$ can be $T^4$ or $K_3$, corresponding to the target space of the sigma model on the CFT side. While in the CFT calculation we focused on the special case of a $T^4$ made from 4 identical $S^1$ factors, the gravity side calculation is completely insensitive to the details of $M_4$. For calculational simplicity, we will only consider the case with a simple self-dual R-R three-form flux where the harmonic functions in \cite{Chiodaroli_2010} only have two singularities, and leave the cases of arbitrary charge deformations to future work (these would correspond to the case of F1-NS5 solutions or more general branes preserving the same amount of supersymmetry). 

To calculate the transmission coefficient $\mathcal{T}$, we need to study a linearized gravity wave propagating on this background, corresponding to a boundary wave incoming from (say) the left, leading to a transmitted and reflected wave due to the presence of the interface.
In \cite{Bachas_2020} this was done for the holographic ICFT dual to a thin-brane model. $\mathcal{T}$ was related to the tension of the thin brane dividing the two parts of the spacetime. Soon after \cite{Baig:2022cnb} showed that the generalization to two thin branes is almost trivial: the tensions simply add. This was generalized to the 'thick-brane' model in \cite{bachas2022energy} where $\mathcal{T}$ of the non-supersymmetric Janus solution was presented. Anytime the geometry only depends non-trivially on a single coordinate, as is in the case of an AdS$_2$ slicing of AdS$_3$, one can approximate the smooth stress tensor corresponding to the fields supporting the geometry by an array of delta functions, effectively treating the spacetime as one with multiple thin branes. In the limit that the comb of delta functions goes back to a smooth configuration one obtains the transmission coefficient of the original geometry\footnote{It would be reassuring to reproduce the formula for $\mathcal{T}$ directly from studying fluctuation around the continuous geometry. This was left for future work in \cite{bachas2022energy}. Here we assume that the answer presented there as one obtains it from taking the continuum limit of the discretized geometry is, in fact, correct.} 

However a caveat with this method is that if there is non-trivial dependence on some extra compact dimensions (like in our super-Janus solution where the metric and matter fields depend non-trivial on both coordinates on $\Sigma$), the method needs to be modified. 
To resolve this issue, in this section we will use Kaluza-Klein compactification to reduce the 10D supergravity above to a 6D supergravity in Einstein frame, and then further  to an effective 3D model, where we can compute the transmission coefficient via the thick-brane model. An alternative way of matching the higher dimensional model with the effective 3D model in the spirit of the above discussion is in Section \ref{sec:fluc} by considering the spin-2 fluctuation wavefunction.

\subsection{6D supergravity with R-R charge}
We consider a type IIB solution on AdS$_2\times S^2\times M_4$ over a strip $\Sigma$.  We will use the most general metric with two asymptotic AdS$_3\times S^3$ regions, which contains five parameters $\theta, \psi, L, k, b$. Following \cite{Chiodaroli_2010}, the relevant functions for our solution are two independent holomorphic functions $A$ and $B$, and two independent harmonic functions $H$ and $\hat{h}$. The complex coordinate on the strip $\Sigma$ will be written as $w=x+iy$ below, where $x$ is non-compact and $y\in [0,\pi]$.

The metric reads
\begin{equation}   ds^2=f_1^2ds_{AdS_2}^2+f_2^2ds_{S^2}^2+f_3^2ds_{M_4}^2+\rho^2ds_{\Sigma}^2,\label{eq:10dmetric}
\end{equation}
where the dilaton and the various functions appearing in the metric are given by 
\begin{equation}
    \begin{split}
        e^{-2\phi}&=\frac{1}{4}\left(A+\bar{A}-\frac{(B+\bar{B})^2}{\hat{h}}\right)\left(A+\bar{A}-\frac{(B-\bar{B})^2}{\hat{h}}\right)\\
        f_3^4&=4\frac{e^{-\phi}\hat{h}}{A+\bar{A}}\\
        f_1^2&=\frac{e^{\phi}}{2f_3^2}\frac{|H|}{\hat{h}}\left((A+\bar{A})\hat{h}-(B-\bar{B})^2\right)\\
        f_2^2&=\frac{e^{\phi}}{2f_3^2}\frac{|H|}{\hat{h}}\left((A+\bar{A})\hat{h}-(B+\bar{B})^2\right)\\
        \rho^4&=e^{-\phi}\hat{h}\frac{|\partial_wH|^4}{H^2}\frac{A+\bar{A}}{|B|^4}\\
    \end{split}
    \label{eq:harmtometric}
\end{equation}

Let us consider the case with nonzero R-R charge and zero NS-NS charge. The generic solutions to the harmonic and holomorphic functions are
\begin{equation}
    \begin{split}
        H&=-iL\sinh(w+\psi)+c.c.\\
        A&=ik^2\frac{\cosh\theta+\sinh\theta\cosh w}{\sinh w}+ib\\
        B&=ik\frac{\cosh(w+\psi)}{\cosh\psi\sinh w}\\
        \hat{h}&=i\frac{\cosh\theta-\sinh\theta\cosh w}{\sinh w}+c.c.\\
    \end{split}
\end{equation}

Under \eqref{eq:harmtometric}, the dilaton is 
\begin{equation}        e^{-2\phi}=k^4\frac{\cosh^2(x+\psi)~\mathrm{sech}^2\psi+(\cosh^2\theta-\mathrm{sech}^2\psi)\sin^2y}{(\cosh x-\cos y\tanh\theta)^2},
\end{equation}
and metric factors are
 \begin{equation}
     \begin{split}
         \rho^4&=e^{-\phi}\frac{L^2}{k^2}\frac{\cosh^2x\cosh^2\theta-\cos^2y\sinh^2\theta}{\cosh^2(x+\psi)}\cosh^4\psi,\\
         f_3^4&=e^{-\phi}\frac{4}{k^2}\frac{\cosh x\cosh\theta-\cos y\sinh\theta}{\cosh x\cosh\theta+\cos y\sinh\theta},\\
         f_1^2&=\rho^2\frac{\cosh^2(x+\psi)}{\cosh^2\theta\cosh^2\psi},\\
         f_2^2&=\rho^2\left(\frac{1}{\sin^2y}+\frac{\cosh^2\theta\cosh^2\psi-1}{\cosh^2(x+\psi)}\right)^{-1}.\\
     \end{split}
 \end{equation}
 
The pure AdS$_3\times S^3$ solution corresponds to $\theta=\psi=0$. At the asymptotic regions of the strip $x\to\pm\infty$ near the ICFT boundary the six dimensional dilaton dual to the volume of $T^4$ is
\begin{equation}
    \lim_{x\to\pm\infty}e^{-\phi}f_3^4=4k^2\mathrm{sech}^2\psi~e^{\mp 2\psi}
\end{equation}

Similarly, the asymptotic values for the combination of the axion $\chi$ and the RR four form $C_4$ near the boundary are
\begin{equation}
    \lim_{x\to\pm\infty}e^{\phi/2}f_3^2\chi-4\frac{e^{-\phi/2}}{f_3^2}  C_4=\pm 4k\sinh\theta
\end{equation}
and is dual to the $\mathbb{Z}_2$ orbifold twist operator $T_0$ mentioned above. From the above two equations it is clear that the parameters $k,\psi,\theta$ are the same ones inside the variable change \eqref{eq:variable}.



\subsection{KK compactify to 3D}

Our goal is to reduce the system to an effective 3D geometry, so we can use the expression in \cite{bachas2022energy} to derive the transmission coefficient for the system on the strip. 

We can consider compactifying the above 10D metric down to 6D, keeping it in the Einstein frame. The 6D metric is AdS$_2\times S^2$ fibred over $\Sigma$, which simplifies to \cite{Chiodaroli_2010}
\begin{equation}
    ds_{6D}^2=R^2~K(x,y)\left(\frac{\cosh^2(x+\psi)}{\cosh^2\psi\cosh^2\theta}ds_{AdS_2}^2+dx^2+dy^2\right)+\frac{R^2\sin^2y}{K(x,y)}ds_{S^2}^2
    \label{eq:metric}
\end{equation}
where $R^2=2L\cosh\psi\cosh\theta$ is the AdS$_3$ radius and
\begin{equation}
    K(x,y)=\sqrt{1+\frac{(\cosh^2\theta\cosh^2\psi-1)\sin^2y}{\cosh^2(x+\psi)}}.
\end{equation}
For the AdS$_2$ metric, we will use the Poincare patch $ds_{AdS_2}^2=(dz^2-dt^2)/z^2$.

The reduction so far was fairly standard as the metric was completely independent of the coordinates on $T^4$. The non-trivial dependence of the overall size of the $T^4$ on $\Sigma$ as encoded in $f_3$ simply got absorbed into the 6D dilaton. To further reduce to 3D, we have to be very careful as, in addition to the $S^2$, we need to reduce on the compact part of $\Sigma$, on which all functions depend. 

We start with the 6D Einstein-Hilbert action $S=1/(16\pi G_6)\int\sqrt{-g_6}R_6$, and integrate out the sphere $S^2$ trivially. This gives us an effective 4D Einstein-Hilbert action with $x,y$-dependent Newton constant $S=1/16\pi\int G_4(x,y)^{-1}\sqrt{-g_4}R_4$.



Now we further compactify along the compact $y$ direction by integrating the above action over $y\in [0,\pi]$. Effectively, we now have a 3D action $S=1/16\pi \int G_3(x)^{-1}\sqrt{-g_3}R_3$ with a $x$-dependent Newton constant $G_3(x)$, and a certain 3D metric that obeys the Einstein equation. After changing the frame for the 3D metric $g_3\to g_3~G_3(x)^{1/2}$, we got the standard Einstein-Hilbert action $S=1/16\pi G_3 \int\sqrt{-g_3}R_3^{EH}$. Explicitly, after integrating out $S^2$ and $y$ and gauging away $G_3(x)$, the action becomes
\begin{equation}
    S=-\frac{R^4}{48\pi G_6}\int_{AdS_2\times\mathbb{R}_x}\frac{3\cosh 2(x+\psi)+\cosh 2\psi+2}{z^2\cosh^2\psi\cosh^2\theta }.
    \label{eq:int6d}
\end{equation}

The ansatz for such a 3D solution is an AdS$_2$-slicing 3D effective model with warpfactor $e^{2A(x)}$:
\begin{equation}
    ds_\text{3D}^2=R^2\left(e^{2A(x)}ds_{AdS_2}^2+dx^2\right).
    \label{eq:warpfactor}
\end{equation}

It has an Einstein-Hilbert action of
\begin{equation}
    S_\text{3D}=-\frac{R}{16\pi G_3}\int\frac{e^{2A}(2e^{-2A}+6A'^2+4A'')}{z^2}.
\end{equation}

Matching the above ansatz with the integrated 6D action from \eqref{eq:int6d} and using the KK relation of Newton constants $G_6=R^3G_3/3$, the solution is
\begin{equation}
    A(x)=\log\left(\frac{\cosh(x+\psi)}{\cosh\psi\cosh\theta}\right),
\end{equation}

and that gives us the effective 3D metric
\begin{equation}
    ds_{\text{3D}}^2=R^2\left(\frac{\cosh^2(x+\psi)}{\cosh^2\psi\cosh^2\theta}ds_{AdS_2}^2+dx^2\right)
    \label{eq:toymodel}
\end{equation}

\subsection{$\mathcal{T}$ from the 3D effective model}
In order to calculate the transmission coefficient from the thick brane method as in \cite{bachas2022energy}, we do a change in coordinate $x$ to a compact $\tilde{x}$, lining up with the ones used in the cited paper. Consider $\tilde{x}\in(-\pi\cosh\psi\cosh\theta/2,\pi\cosh\psi\cosh\theta/2)$ where
\begin{equation}
    \cos\frac{\tilde{x}}{\cosh\psi\cosh\theta}=\frac{1}{\cosh(x+\psi)}
\end{equation}



The effective 3D model above then becomes
\begin{equation}
    ds_{\text{toy}}^2=R^2e^{2A(\tilde{x})}\left(ds_{AdS_2}^2+d\tilde{x}^2\right)
\end{equation}

where
\begin{equation}
    e^{2A(\tilde{x})}=\left(\cosh^2\psi\cosh^2\theta~\cos^2\frac{\tilde{x}}{\cosh\psi\cosh\theta}\right)^{-1}.
\end{equation}

From Eq. (2) and Eq. (17) in \cite{bachas2022energy}, we finally obtain the strong coupling transmission coefficient of the super-Janus solution
\begin{equation}
    \mathcal{T}_{\text{strong}}=\frac{2c_{LR}}{c_L+c_R}=\frac{1}{\cosh^2\psi\cosh^2\theta}
    \label{eq:transmission}
\end{equation}
which, as advertised, is identical to the free field theory answer of \eqref{eq:tweak} with jumping radii and $B$ field deformation.

\section{Alternative Derivation: Graviton reduction for AdS$_2$/CFT$_1$}
\label{sec:fluc}
Another way to derive the transmission coefficient is via graviton fluctuation as illustrated in Section 5 of \cite{bachas2022energy}. Previously,  the reduction of the AdS graviton from a higher-dimensional solution has been done in various dimensions \cite{Bachas_2011,Speziali_2020,Chen_2019}. Below, we will derive the fluctuation wavefunction $\Psi$ for the AdS$_2$ graviton from the 10D solution, and will indirectly read off $\mathcal{T}$ from the asymptotic behavior of $\Psi$ at the ends of the strip. 

\subsection{Fluctuation wavefunction in AdS$_2$/CFT$_1$}
We wish to obtain the spin-2 spectrum reduced from a top-down supergravity solution to AdS$_2$. Following \cite{Rigatos_2023}, we rewrite the perturbations of the 10D metric \eqref{eq:10dmetric} as
\begin{equation}
    ds^2=e^{2A(w)}(\bar{g}_{\mu\nu}+h_{\mu\nu})dX^\mu dX^\nu+\hat{g}_{ab}dw^adw^b,
\end{equation}

where $X^\mu$ are the AdS$_2$ coordinates, $w^a$ are for the internal manifold, and $e^{2A}$ is the warpfactor. We want to find the AdS$_2$ graviton fluctuation of the form 
\begin{equation}
    h_{\mu\nu}(X,u)=h_{\mu\nu}^{[tt]}(X|\lambda)\Psi(w|\lambda)
\end{equation}

where $h_{\mu\nu}^{[tt]}(X|\lambda)$ is the Pauli-Fierz transverse-traceless solution for a spin-2 particle in AdS with mass $m^2=\lambda+2$. For our 10D supergravity, the internal wavefunction $\Psi(w|\lambda)$ is solved by \cite{Bachas_2011,Rigatos_2023}
\begin{equation}
    -\frac{1}{\sqrt{[\hat{g}]}}(\partial_a\sqrt{[\hat{g}]}\hat{g}^{ab}e^{2A}\partial_b)\Psi=m^2\Psi,
    \label{eq:dhoker}
\end{equation}

where $[\hat{g}]$ is the determinant of $\hat{g}_{ab}$.\\

 We can see that the lowest mass solution has $l=0$ and the lowest energy $\Delta_4=0$ for $M_4$. 

In our metric the above $\hat{g}$ is 
\begin{equation}    \hat{g}_{ab}dy^ady^b=f_2^2ds_{S^2}^2+f_3^2ds_{M_4}^2+\rho^2dzd\bar{z},
\end{equation}

and we will use the spherical harmonic function for $\Psi$ as
\begin{equation}
    \Psi=Y_{lm}\Psi_l(x,y).
\end{equation}

Plugging this ansatz into \eqref{eq:dhoker}, we have
\begin{equation}
    -f_1^2\left(\frac{\delta^{ij}}{\rho^2}\partial_i\partial_j+\frac{2\delta^{ij}}{\rho^2}(\partial_i\log(f_1f_2f_3^2))\partial_j-\frac{l(l+1)}{f_2^2}-\frac{\Delta_4}{f_3^2}\right)\Psi_l(y^1,y^2)=m^2\Psi_l(y^1,y^2).
    \label{eq:pde}
\end{equation}

Since we will only consider $\Delta_4=l=0$, below we abuse the notation and use $\Psi$ for the 0th component $\Psi_0$. We can substitute $\Psi$ with a more convenient $\tilde{\Psi}$:
\begin{equation}
    \tilde{\Psi}=\frac{\cosh(x+\psi)\sin y}{2\cosh\psi\cosh\theta}\Psi, 
\end{equation}

and write the equation for the $l=0$, $\Delta_4=0$ component as
\begin{equation}
    -\cosh^2(x+\psi)(\partial_x^2+\partial_y^2)\tilde{\Psi}=m^2\cosh^2\psi\cosh^2\theta~\tilde{\Psi}.\\
\end{equation}

The variables are separable in this case, and imposing the Neumann boundary condition on $y$ for $\Psi$, we have the general solution $\tilde{\Psi}_0=\sum_n\tilde{\xi}_n(x)\sin((2n+1)y)$, where $\tilde{\xi}_n$ satisfies the Legendre equation, and has solution
\begin{equation}
    \tilde{\xi}_n=c_1~P_\kappa^{2n+1}(\tanh (x+\psi))+c_2~Q_\kappa^{2n+1}(\tanh (x+\psi)),~~\kappa(\kappa+1)=m^2\cosh^2\psi\cosh^2\theta.
\end{equation}

For a nonzero and convergent $\tilde{\xi}_n$, $\kappa$ has to be an integer greater than 0, and $c_2=0$. At first glance, the lowest massive excitation of AdS$_2$ is $m^2=2/(\cosh^2\psi\cosh^2\theta)$.

However, the $m^2=0$ excitation corresponds to the wavefunction satisfying
\begin{equation}
    (\partial_x^2+\partial_y^2)\tilde{\Psi}=0,
\end{equation}

which after seperating variables gives a nontrivial equation
\begin{equation}    \tilde{\Psi}=\sum_n(A_n\cosh(2n+1)(x+\psi)+B_n\sinh(2n+1)(x+\psi))\sin(2n+1)y.
\end{equation}

The non-divergent terms for $\Psi$ are then only the $n=0$ components:
\begin{equation}
    \Psi=A_0+B_0\tanh (x+\psi).
    \label{eq:fluc}
\end{equation}

In principle we should be able to extract $\mathcal{T}$ from this fluctuation.

\subsection{From $\Psi$ to $\mathcal{T}$}

The fluctuation wavefunction \eqref{eq:fluc} in principle encodes all the information about the fluctuation and one should be able to directly read off $\mathcal{T}$ from it. A general recipe of how one would go about doing this has been outlined in \cite{bachas2022energy}, even though the calculation hasn't been carried out explicitly as one has to deal with matching the various asymptotic forms of the metric which proves to be surprisingly cumbersome.

As a workaround, we are going to pursue the following alternate strategy. As the transmission coefficient is encoded in the graviton fluctuation wavefunction \eqref{eq:fluc}, we can simply construct a toy model in 3D that has the same fluctuation function. Explicitly, the PDE \eqref{eq:pde} on $x,y$ for the wavefunction $\Psi(x,y)$ with mass zero is
\begin{equation}
    \frac{\cosh^2(x+\psi)}{\cosh^2\psi\cosh^2\theta}\left(2\tanh(x+\psi)~\partial_x+\partial_x^2+2\cot y~\partial_y+\partial_y^2\right)\Psi(x,y)=0.
\end{equation}

Its solution \eqref{eq:fluc} is actually the solution to an ODE eigenvalue equation on $x$. Namely, let us reduce the $y$ direction of the above equation and consider the fluctuation of a 3d gravity
\begin{equation}
\left(\partial_x^2+2\tanh(x+\psi)~\partial_x+M^2\right)\Psi(x)=0.
    \label{eq:ode}
\end{equation}

Recall the ansatz for 3D Einstein gravity with isometry $SO(2,1)$ in the form of \eqref{eq:warpfactor}. Analogous to the modeling in \cite{DeWolfe_2000} for AdS$_4$-slicing of a 5d model, the fluctuation wavefunction of the transverse-traceless mode $\Psi(x)$ of 3D gravity obeys the ODE
\begin{equation}
    \left(\partial_x^2+2A'(x)\partial_x-e^{-2A}\Box\right)\Psi(x)=0
    \label{eq:3dfluc}
\end{equation}

Comparing \eqref{eq:fluc} with \eqref{eq:ode}, the warpfactor $A(x)$ satisfies
\begin{equation}
    A'(x)=\tanh(x+\psi).
\end{equation}
Integrating the above equation we get the warpfactor $e^{2A}=k\cosh^2(x+\psi)$. To determine the constant $k$ we demand that the toy model has the same asymptotic behavior in the $x\to\pm\infty$ region as the original metric \eqref{eq:metric}. Low and behold we reach a 3D toy model with a metric that is identical to what we found by explicitly doing the KK reduction  in \eqref{eq:toymodel}. 

Strictly speaking, this alternate method can be viewed as a check of the KK-reduction procedure in the previous chapter: we explicitly verified that the fluctuation equation in the full 6D metric is identical to the one of the effective 3D metric. 

As before, we now can simply plug in the 3D toy model metric into the formulas of \cite{bachas2022energy} and, of course, once more obtain \eqref{eq:tweak}.

More generally, this procedure allows us to derive the transmission coefficient $\mathcal{T}$ from the graviton fluctuation wavefunction, in particular for top-down models with internal dimensions dependence where the spin-2 wavefunction happens to be separable.


\newpage


\section{Discussion and future directions}
\label{sec:discussion}

In this work we demonstrated that the remarkable agreement between strongly coupled super-Janus that was previously seen in $c_\text{eff}$ and $g$ continues to hold for the transmission coefficient. The obvious next future direction is understanding why. Clearly novel non-renormalization theorems are at work in these supersymmetric ICFTs.

Another interesting finding involves the two important quantities in an (1+1)D ICFT: entanglement entropy $c_\text{eff}$ and transmission coefficient $c_{LR}$. They are interestingly bounded by each other \cite{Karch:2024udk}. According to the calculations in \cite{Gutperle:2015hcv}, the entanglement entropy across the interface in the super Janus solution in Section \ref{sec:holography} is 
\begin{equation}
    c_\text{eff}=\frac{c}{\cosh\psi\cosh\theta}
\end{equation}

in the strong-coupling gravity calculation, and agrees with the weak-coupling limit of the dual 2d $\mathcal{N}=(4,4)$ orbifold ICFT\footnote{The weak-coupling CFT calculation for the entanglement entropy was given under no jumping RR modulus, i.e. $\theta=0$, and was not yet performed in presence of the orbifold twist $\theta$.}, just like the transmission coefficient. Morever, we observe that in presence of supersymmetry, the relation between $c_\text{eff}$ and $c_{LR}$ becomes more stringent: in the case of super Janus solution where $c_{LR}$ is given by \eqref{eq:transmission} and $c_L=c_R=c$, we observe a simple relation 
\begin{equation}
    \mathcal{T}=\frac{c_{LR}}{c}=\left(\frac{c_\text{eff}}{c}\right)^2.
    \label{eq:relation}
\end{equation}

One can ask whether this relation holds true for most general type IIB supergravity BPS solutions in the form of \eqref{eq:10dmetric} with two poles in the harmonic functions corresponding to two asymptotic AdS$_3$ regions. Concretely, from the BPS equations for the metric factors $f_1^2f_2^2f_3^4=H^2$ we can write the general 6D metric in terms of the harmonic functions $H,\hat{h}$ and holomorphic functions $A,B$ as
\begin{equation}
\begin{split}
    ds^2=\rho^2f_3^2&\left(\frac{H^2|B|^2}{|\partial_wH|^2((A+\bar{A})\hat{h}-(B+\bar{B})^2)}ds_{AdS_2}^2+dx^2+dy^2\right)\\
    &+\frac{|\partial_wH|^2((A+\bar{A})\hat{h}-(B+\bar{B})^2)}{|B|^2\rho^2f_3^2}ds_{S^2}^2
\end{split}  
\end{equation}

and the KK reduced 3D action after integrating the $y$ direction is expected to have the 3D effective model in the form of
\begin{equation}
    ds_\text{3D}^2=R^2(k^2\cosh^2(x+s)~d_{AdS_2}^2+dx^2)
\end{equation}

which leads to the above relation between $c_\text{eff}$ and $c_{LR}$. In fact, this relation has already been pointed out for supersymmetric free theories with $c=3/2$ across a conformal interface \cite{Brehm:2015lja}. It is curious to see whether it is a general rule for supersymmetric ICFT with a conformal interface.

Another interesting question is to further explore the role of the internal manifold. Our supergravity answer did not at all depend on the shape of the internal torus, or even whether we replace $T^4$ with K3 altogether. On the field theory side we limited ourselves to a very symmetric torus. The supergravity answer makes one suspect that in the field theory $\mathcal{T}$ should also be independent of the shape of the internal manifold. It would be interesting to verify this explicitly.

\section*{Acknowledgements}

We'd like to thank Costas Bachas for collaboration at initial stages of this work, and him as well as Stefano Baiguerra, Shira Chapman, Giuseppe Policastro, and Tal Schwartzman for discussions of their work in \cite{bachas2022energy}. Thanks to Ilka Brunner, Michael Gutperle, Yuya Kusuki, and Hiroshi Ooguri for comments on a draft of our manuscript.
This work was supported in part by the U.S. Department of Energy under Grant No. DE-SC0022021 and a grant from the Simons Foundation (Grant 651678, AK). 

\bibliographystyle{JHEP}
\bibliography{transmission,references}

\end{document}